\documentclass[aps,prd,preprint,superscriptaddress,showpacs,floatfix,%
nofootinbib]{revtex4}
\usepackage{epsfig}
\usepackage{latexsym}
\usepackage{bm}
\usepackage{amsmath}
\usepackage{colordvi}

\begin{document}
 
\preprint{\rightline{ANL-HEP-PR-05-9}}

\title{Spin correlations and velocity-scaling in color-octet NRQCD matrix 
elements}

\author{Geoffrey T.~Bodwin}
\affiliation{
High Energy Physics Division,
Argonne National Laboratory,
9700 South Cass Avenue, Argonne, Illinois 60439}
 
\author{Jungil Lee}
\affiliation{
Department of Physics,
Korea University,
Seoul 136-701, Korea}

\author{D.K.~Sinclair}
\affiliation{
High Energy Physics Division,
Argonne National Laboratory,
9700 South Cass Avenue, Argonne, Illinois 60439}

\begin{abstract}
We compute spin-dependent decay matrix elements for $S$-wave
charmonium and bottomonium in lattice nonrelativistic quantum
chromodynamics (NRQCD). Particular emphasis is placed upon the
color-octet matrix elements, since the corresponding production matrix
elements are expected to appear in the dominant contributions to the
production cross sections at large transverse momenta. We use three
slightly different versions of the heavy-quark lattice Green's functions
in order to minimize the contributions that scale as powers of the
ultraviolet cutoff. The lattice matrix elements that we calculate obey
the hierarchy that is suggested by the velocity-scaling rules of
NRQCD.
\end{abstract}


\maketitle

\section{Introduction}

The production rates of the $J/\psi$ and the $\Upsilon$ at large
transverse momentum ($p_T$) at the Fermilab Tevatron provide important
tests of our understanding of heavy-quarkonium systems and of quantum
chromodynamics (QCD) itself. The dominant mechanism for both $J/\psi$
and $\Upsilon$ production at large $p_T$ is expected to be the
fragmentation of a gluon into a heavy-quark ($Q\bar{Q}$) pair
\cite{Braaten:1993rw,Braaten:1995cj}. Because the heavy-quark mass $m$
provides a large momentum scale, in addition to $p_T$, the production of
this heavy-quark pair can be calculated perturbatively. The evolution of
the $Q\bar{Q}$ pair into the $J/\psi$ or the $\Upsilon$ is a
nonperturbative process. In the color-singlet model
\cite{Einhorn:1975ua,Ellis:1976fj,Carlson:1976cd,Kuhn:1979kb,
DeGrand:wf,Kuhn:1979zb,Wise:1979tp,Chang:1979nn,Berger:1980ni,Baier:1981zz,
Baier:1981uk,Keung:1981gs,Baier:1983va}, it is assumed that the $Q\bar
Q$ pair is in a color-singlet state. That assumption leads to
predictions that are more than an order of magnitude below the Tevatron
data \cite{Brambilla:2004wf}. A less {\it ad hoc} approach to quarkonium
production is based on the effective field theory nonrelativistic QCD
(NRQCD) \cite{Caswell:1985ui,Thacker:1990bm,Bodwin:1994jh}. This approach
is known as NRQCD factorization. The predictions of NRQCD factorization
fit the Tevatron data and indicate that the dominant production
mechanism at large $p_T$ proceeds through the fragmentation of a gluon
into a $Q\bar Q$ pair in a color-octet state.

The effective field theory NRQCD separates long-distance,
nonperturbative quarkonium dynamics ($p \lesssim mv$) from
short-distance, perturbatively calculable processes ($p \gtrsim m$).
Here $m$ is the heavy-quark mass, $p$ is the magnitude of the relative
three-momentum of the $Q$ and $\bar Q$ in the quarkonium rest frame, and
$v$ is the typical relative velocity of the $Q$ and $\bar Q$ in the
quarkonium rest frame. If $\Lambda$ is the ultraviolet (UV) cutoff of
NRQCD, then physics with $p < \Lambda \sim m$ is reproduced in the
effective field theory, while physics with $p > \Lambda$ is integrated
out, but affects the coefficients of local interactions in the effective
theory. The Lagrangian of NRQCD can be expanded in powers of the
velocity $v$. To any finite order in $v$ only a limited number of
interactions appears in this Lagrangian. Hence, it is useful in
calculations for systems in which $v^2 \ll 1$. We note that in
bottomonium $v^2 \approx 0.1$, while in charmonium $v^2 \approx 0.3$.

In the context of NRQCD, scaling rules can be deduced for the leading
behavior of operator matrix elements in the limit in which the
heavy-quark velocity $v$ approaches zero
\cite{Lepage:1992tx,Bodwin:1994jh}. When these ``$v$-scaling rules'' are
applied to the production of the $J/\psi$ and the $\Upsilon$ at large
$p_T$, they predict that, in the nonperturbative evolution of the $Q\bar
Q$ pair into the quarkonium, the non-spin-flip interactions dominate over
the spin-flip interactions, with corrections of order $v^2$. Hence, the
$J/\psi$ is predicted to take on most of the transverse polarization of
the gluon\cite{Cho:1994ih}. However, the CDF data for the polarization
\cite{Affolder:2000nn} show decreasing transverse polarization with
increasing $p_T$ and disagree with the NRQCD prediction
\cite{Braaten:1999qk} in the largest $p_T$ bin.

The existing calculations of polarization at the Tevatron neglect
spin-flip processes in the NRQCD matrix elements, under the assumption
that, because of the $v$-scaling rules, the spin-flip contributions are
relatively suppressed. However, the $v$-scaling rules predict only the
leading power of $v$, not its coefficient. It is usually assumed in
making estimates that the coefficients are of order unity, but it could
happen that the coefficients of the spin-flip terms are anomalously large
or that the coefficients of the non-spin-flip terms are anomalously
small. It has also been suggested that the $v$-scaling rules
themselves may need to be modified in the case of charmonium
\cite{Beneke:1997av,Brambilla:1999xf,Fleming:2000ib,Sanchis-Lozano:2001rr,
Brambilla:2002nu}.

One would like to test the applicability of estimates based on
the $v$-scaling rules by direct calculation of the relevant NRQCD
operator matrix elements. At present, lattice QCD is the only technique
that is available for calculating color-octet matrix elements.
Unfortunately, it is not known how to formulate that calculation of
color-octet production matrix elements in Euclidean lattice field
theory. However, one does know how to calculate the corresponding decay
matrix elements in lattice NRQCD \cite{Thacker:1990bm,Davies:1994mp,bks}.
Since the decay matrix elements are predicted to obey the same
velocity-scaling rules as their production counterparts, a test of
$v$-scaling estimates in the context of decay matrix elements might shed
some light on the validity of $v$-scaling estimates for production
matrix elements. In this paper, we present lattice NRQCD calculations of
the decay matrix elements that are related by crossing to those
production matrix elements that are expected to dominate $S$-wave
quarkonium production at large $p_T$. Preliminary results of these
calculations were presented in Ref.~\cite{Bodwin:2004up}.

We use the methods developed by Lepage {\it et al.} \cite{Lepage:1992tx}
to compute heavy-quark Green's functions in lattice NRQCD. From these, we
construct quarkonium propagators and use the methods of
Refs.~\cite{Davies:1994mp,bks} to extract the spin-dependent color-octet
matrix elements that contribute to the decays of the $J/\psi$, $\eta_c$,
$\Upsilon$, and $\eta_b$ ($S$-wave quarkonia) in lattice NRQCD. We
perform the lattice measurements on a set of 400 $12^3 \times 24$
quenched configurations at $\beta=6/g^2=5.7$. This $\beta$ value was
chosen because it corresponds to a momentum space cutoff $\Lambda \sim
m_b$. In the case of charmonium one would like to have $\Lambda \sim
m_c$. However, for lattices that are significantly coarser than those
with $\beta=5.7$, it is not clear, even with highly improved actions,
that one is close enough to the continuum limit to make contact with
perturbation theory. We therefore use $\beta=5.7$ lattices for
charmonium and study some of the effects of those contributions that
diverge as powers of $\Lambda/m_c$ by varying the algorithm for
calculating lattice heavy-quark Green's functions.

Our bottomonium measurements suggest that the NRQCD $v$-scaling rules are
useful in determining which matrix elements are most important, and that
those contributions that are suppressed as powers of $v$ show an
even greater suppression than one would expect from the $v$-scaling
factors alone. For charmonium the situation is complicated by our
inability to use $\Lambda \sim m_c$, but there are indications that the
$v$-scaling rules are also a good guide here. Calculation of the
perturbative coefficients that relate lattice and continuum matrix
elements could help to clarify this situation.

In Sec.~\ref{sec:NRQCD} we introduce the NRQCD 
Lagrangian through relative order $v^4$, NRQCD factorization of
quarkonium production and decay rates, and the lattice implementations
of NRQCD that we use. We discuss the calculations that we performed
and present our results in Sec.~\ref{sec:calcs-results}.
Sec.~\ref{sec:summary} contains discussions and our conclusions.

\section{NRQCD} 
\label{sec:NRQCD}
\subsection{Continuum NRQCD}

As we indicated in the Introduction, NRQCD is an effective field theory
with a UV momentum-space cutoff $\Lambda \sim m$. It is useful in
describing bound states of heavy quarks. In the case of $Q\bar Q$ bound
states (quarkonium), the terms in the effective Lagrangian can be
classified according to their leading power behavior in $v$, where $v$
is the typical heavy-quark (or antiquark) velocity in the quarkonium rest
frame \cite{Thacker:1990bm,Lepage:1992tx}. The terms of leading order in
$v^2$ in the NRQCD Lagrangian density are just the Schr\"{o}dinger
Lagrangian density:
\begin{equation}
{\cal L}_0
= \psi^\dagger  \left( iD_t + \frac{{\bf D}^2}{2m} \right) \psi
+ \chi^\dagger  \left( iD_t - \frac{{\bf D}^2}{2m} \right) \chi,
\end{equation}
where $D_t=\partial_t+igA_0$, ${\bf D}=\bm{\partial}-ig{\bf A}$, $\psi$
is the Pauli spinor field that annihilates a heavy quark, and $\chi$ is
the Pauli spinor field that creates a heavy antiquark.
 
In order to reproduce QCD completely, we would need an infinite number of
interactions.  For example, at next-to-leading order in $v^2$ we have
\begin{eqnarray}
\delta{\cal L}_{\rm bilinear}&=& \frac{c_1}{8m^3}
\left[ \psi^\dagger ({\bf D}^2)^2 \psi \;-\; \chi^\dagger ({\bf D}^2)^2
\chi
\right]  \nonumber \\
&+& \frac{c_2}{8m^2}
\left[ \psi^\dagger ({\bf D} \cdot g {\bf E} - g {\bf E} \cdot {\bf D})
\psi
+\chi^\dagger ({\bf D} \cdot g {\bf E} - g {\bf E} \cdot {\bf D}) \chi
\right] \nonumber \\
&+& \frac{c_3}{8m^2}
\left[ \psi^\dagger (i {\bf D} \times g {\bf E} - g {\bf E} \times i
{\bf D})
\cdot \bm{\sigma} \psi
+\chi^\dagger (i {\bf D} \times g {\bf E} - g {\bf E} \times i {\bf D})
        \cdot \bm{\sigma} \chi \right] \nonumber \\
&+& \frac{c_4}{2m}
\left[ \psi^\dagger (g {\bf B} \cdot \bm{\sigma}) \psi
        \;-\; \chi^\dagger (g {\bf B} \cdot \bm{\sigma})
\chi \right].
\label{lbilinear}
\end{eqnarray}
In practice, we work to a given precision in $v$. For the calculations
presented in this paper, the contributions given above suffice.

It has been conjectured that, at large transverse momentum $p_T$, the
inclusive quarkonium production cross section can be written in a
factorized form
(Ref.~\cite{Bodwin:1994jh}):
\begin{equation}
\sigma(H)=\sum_n \frac{F_n(\Lambda)}{m^{d_n-4}}\langle 0|
{\cal O}_n^H(\Lambda)|0\rangle.
\label{prod-fact}
\end{equation}
The ``short-distance coefficients'' $F_n(\Lambda)$ are essentially the
partonic cross sections to make a $Q\bar Q$ pair with a given set of
quantum numbers convolved with parton distributions. These partonic
cross sections can be calculated as an expansion in $\alpha_s$. The
$F_n$ multiply vacuum matrix elements of four-fermion operators of the
form
\begin{equation}
{\cal O}_n^H=\chi^\dagger \kappa_n\psi
\left(\sum_X |H+X\rangle\langle H+X|\right) \psi^\dagger
\kappa'_n\chi.
\label{matrix-element-defn}
\end{equation}
$\kappa$ contains Pauli matrices, color matrices, and the covariant
derivatives $D_t=\partial_t+igA_0$, ${\bf D}=\bm{\partial}-ig{\bf
A}$.\footnote{A recent study of certain two-loop contributions to
quarkonium production \cite{Nayak:2005rw} has revealed that, if
factorization is to hold, then the matrix elements must be modified from
the form given in Eq.~(\ref{matrix-element-defn}) by the inclusion of
lightlike eikonal lines that run from each of the $Q\bar Q$ bilinears
to the far future. It is not known if this modification preserves the
factorized form in higher orders.}
The operator matrix elements contain all of the long-distance,
nonperturbative physics and are, essentially, the probabilities for a
$Q\bar Q$ pair with a given set of quantum numbers to evolve into a 
heavy-quark\-o\-ni\-um state. The matrix elements are universal, {\it i.e.,}
process independent. NRQCD predicts the leading scaling behavior of the
matrix elements with $v$ (Ref.~\cite{Bodwin:1994jh}). As a consequence of these
$v$-scaling rules, the sum over operator matrix elements can be regarded
as an expansion in powers of $v$, where $v^2\approx 0.3$ for charmonium and
$v^2\approx 0.1$ for bottomonium. 

A similar factorization formula applies to inclusive quarkonium decays
\cite{Bodwin:1994jh}:
\begin{equation}
\Gamma(H \to {\rm LH})
= \sum_n \frac{2  {\rm Im } f_n(\Lambda)}{m_Q^{d_n-4}} 
        {\langle H |} {\cal O}_n(\Lambda) {| H \rangle},
\label{decay-fact}
\end{equation}
except that the matrix elements are now between quarkonium states, rather than
vacuum states, and the four-fermion operators have the form 
${\cal O}_n=\psi^\dagger \kappa_n\chi\chi^\dagger\kappa_n'\psi$. While the
coefficients $f_n(\Lambda)$ can be calculated perturbatively, the quarkonium
matrix elements ${\langle H |} {\cal O}_n(\Lambda) {| H \rangle}$ are
nonperturbative and can be calculated directly in lattice NRQCD. An
important feature of NRQCD factorization is that both quarkonium decay and
production occur through color-octet, as well as color-singlet, $Q\bar Q$
states. While the production matrix elements are the crossed versions of
quarkonium decay matrix elements, only the color-singlet production and decay
matrix elements are simply related.

The NRQCD operator matrix elements in Eqs.~(\ref{prod-fact}) and
(\ref{decay-fact}) depend explicitly on the cutoff $\Lambda$. In the
physical decay and production rates, this cutoff dependence in the
matrix elements is canceled by a corresponding cutoff dependence in the
short-distance coefficients. However, we note that the $v$-scaling rules for
NRQCD matrix elements are derived under that assumption that the UV
cutoff $\Lambda$ is of order $mv$. The matrix elements of NRQCD, in
common with those of other effective field theories, contain
contributions that diverge, in the limit $\Lambda \rightarrow \infty$,
both as logarithms and powers of $\Lambda/m$. When $\Lambda$ is larger
than $mv$, these divergent contributions potentially spoil the
$v$-scaling rules, with the power divergences being especially important
numerically. 

As an effective field theory, NRQCD is expected to be valid up to values
of $\Lambda$ close to $m$ \cite{Bodwin:1994jh,Lepage:1992tx}. For
bottomonium, $\Lambda \approx m_b$ is large enough that the
physics at momenta greater than $\Lambda$ can be treated perturbatively.
While this choice should not invalidate the $v$-scaling rules
for UV convergent contributions to matrix elements, it remains to be seen
whether the $v$-scaling rules remain valid for matrix elements whose
leading contributions diverge as $\Lambda \rightarrow \infty$. The
measurements described in this paper test this conjecture, as well 
as the $v$-scaling rules for $\Lambda\sim mv$. In
lattice evaluations of the NRQCD matrix elements, the value of the
cutoff is determined, in part, by the lattice spacing. As we shall
see, the effective cutoff in the matrix elements is also affected by the
specific forms of the lattice action and Green's functions that are
employed.\footnote{In the continuum phenomenology of quarkonium
production, the short-distance coefficients are usually calculated in
dimensional regularization. In dimensional regularization, power
infrared divergences in the short-distance coefficients are set to zero.
This implies that, in the corresponding continuum NRQCD matrix elements,
the UV power-divergent contributions are removed order-by-order in
perturbation theory. Therefore, if the lattice matrix elements are to be
close in value to the continuum matrix elements, then the effective
$\Lambda$ in the lattice calculations must be of order the heavy-quark
mass or less.} This latter property allows one to control the
effective cutoff in NRQCD matrix elements without affecting the
interactions that determine the quarkonium masses and wave functions.
Such an approach is especially useful in the case of charmonium, for
which the cutoff $\Lambda=m_c$ is too small to include all of the
nonperturbative bound-state physics.

\subsection{Lattice NRQCD}

The first part of this subsection summarizes the lattice formulation of 
NRQCD that has been given by Lepage {\it et al.}
\cite{Lepage:1992tx}. 

In order to produce a lattice formulation of NRQCD, one must first
formulate NRQCD in Euclidean space. This can be accomplished by
performing the following substitutions in the NRQCD Lagrangian:
\begin{eqnarray}
t            &\rightarrow&    -i t,              \nonumber \\
\partial_t   &\rightarrow&    i \partial_t,      \nonumber \\
D_t          &\rightarrow&    i D_t,             \nonumber \\
\phi         &\rightarrow&    -i \phi,           \nonumber \\
{\bf E}      &\rightarrow&    -i {\bf E}.        
\end{eqnarray}
\label{mink-eucl}

The gauge fields are incorporated into unitary matrices $U_\mu(x)$, which, 
as usual, are defined on the links of the lattice. Covariant derivatives 
are replaced on the lattice by covariant finite differences,
which are defined by
\begin{eqnarray}
\Delta_i \psi(x) &\equiv& \frac{1}{2} [ U_i(x) \psi(x+\hat{\imath}) 
          -U^{\dag}_i(x-\hat{\imath}) \psi(x-\hat{\imath}) ],   \nonumber \\
\Delta_i^{(2)} \psi(x) &\equiv& \frac{1}{2} [ U_i (x) \psi(x+\hat{\imath}) 
                - 2\psi(x)
                +U^{\dag}_i(x-\hat{\imath}) \psi(x-\hat{\imath}) ], \nonumber \\
\Delta^{(2)} &\equiv& \sum_i \Delta_i^{(2)},                    \nonumber \\
\Delta^{(4)} &\equiv& \sum_i 
\left(\Delta_i^{(2)}\right)^2. 
\end{eqnarray}
Here we have adopted the standard lattice convention of working in a system of
units in which the lattice spacing $a$ has been set to unity. 

Tadpole improvement is implemented by making the replacement
\begin{equation}
U_\mu(x) \rightarrow \frac{ U_\mu(x)}{u_0 },
\label{tadpole-imp}
\end{equation}
with $u_0$ chosen as the fourth root of the average plaquette 
\cite{Lepage:1992xa}. For the tadpole-improved 
action, we expect the perturbation series for quantities at the scale 
$\Lambda$ to converge well \cite{Lepage:1992xa}. Therefore, we 
replace each of the coefficients $c_i$ in $\delta{\cal L}_{\rm
bilinear}$ with its lowest-order value, namely, unity.

The order-$v^2$ lattice Hamiltonian is now
\begin{equation}
H_0 = -\frac{\Delta^{(2)}}{2 m}.
\end{equation}
The order-$v^4$ corrections to this Hamiltonian are
\begin{eqnarray}
\delta H &=& -\frac{(\Delta^{(2)})^2}{8 m^3}
             +\frac{ig}{8 m^2} (\bm{\Delta}\cdot {\bf E} 
               - {\bf E}\cdot \bm{\Delta}) \nonumber \\
         & & -\frac{g}{8 m^2} \bm{\sigma}\cdot (\bm{\Delta}\times {\bf E}
                - {\bf E}\times\bm{\Delta})
             -\frac{g}{2 m} \bm{\sigma}\cdot {\bf B}\nonumber \\
         & & +\frac{\Delta^{(4)}}{24 m} 
             -\frac{\left(\Delta^{(2)}\right)^2}{16 n m^2}.
\label{delta-h}
\end{eqnarray}
All of the terms except the last two are simple discretizations of those 
in the continuum expression for $\delta{\cal L}_{\rm bilinear}$ 
[Eq.~(\ref{lbilinear})]. The second-to-last term is the order-$a^2$ 
correction to the discretization of the ${\bf D}^2$ operator in $H_0$. The
last term is the order-$a^2$ correction to the approximation
$\exp(-H_0)\approx [1-H_0/(2n)]^{2n}$, which is used below in
computing the evolution of a heavy-quark Green's function over one
lattice time step. Here, $n$ is an integer to be specified below. We
note that the form of $\delta H$ in Eq.~(\ref{delta-h}) is not 
tadpole-improved 
correctly by the replacement (\ref{tadpole-imp}). That is
because, in the higher-order derivatives $\Delta^{(4)}$ and
$\left(\Delta^{(2)}\right)^2$, there are canceling factors $U^\dagger
U=1$ that are, incorrectly, replaced with $1/u_0^2$. The corrections to
this replacement amount to constant shifts of the Hamiltonian.
Nevertheless, we retain the form of $\delta H$ in Eq.~(\ref{delta-h}),
with the replacement (\ref{tadpole-imp}), because, as noted by the NRQCD
collaboration, it makes the contributions from $\delta H$ to the
evolution of the heavy-quark Green's functions small. The
smallness of the $\delta H$ contributions can be discerned from the fact
that the spin-averaged ``masses'' of the quarkonia are little
affected by the inclusion of these order-$v^4$ corrections.

We make use of two different forms of the heavy-quark Green's functions,
which are equivalent through order $v^4$. One is the form that was used
in the early spectroscopy papers of the NRQCD collaboration
\cite{Davies:1994mp,Davies:1995db}.  In this form, for a source
$S(\bm{x})\delta_{t,0}$, the retarded Green's function $G_r(\bm{x},t)$
at positive time is given recursively by
\begin{eqnarray}
G_r(\bm{x},0)   &=& S(\bm{x})\delta_{t,0},                          \nonumber \\
G_r(\bm{x},1)   &=& \left(1-\frac{H_0}{2n}\right)^n U_4^{\dag}
                    \left(1-\frac{H_0}{2n}\right)^n G_r(\bm{x},0),  \nonumber \\
G_r(\bm{x},t+1) &=& \left(1-\frac{H_0}{2n}\right)^n U_4^{\dag}
                    \left(1-\frac{H_0}{2n}\right)^n (1-\delta H) G_r(\bm{x},t).
\end{eqnarray}
We call this form of the Green's function the ``nrqcd scheme.'' A
second form of the Green's function is defined for positive time by
\begin{eqnarray}                                                                
G_r(\bm{x},0)   &=& S(\bm{x})\delta_{t,0},                        \nonumber \\
G_r(\bm{x},t+1) &=&\left(1-\frac{\delta H}{2}\right)
                   \left(1-\frac{H_0}{2n}\right)^n U_4^{\dag}                  
                   \left(1-\frac{H_0}{2n}\right)^n 
                   \left(1-\frac{\delta H}{2}\right) G_r(\bm{x},t).     
\end{eqnarray}
This second form was used by the NRQCD collaboration in their heavy-light
meson calculations \cite{Collins:2000ix}. For our calculation, we
modify it with  the rule that we replace any factor $\left(1-\delta
H/2\right)^2$ that appears as a result of consecutive time-evolution
steps with $(1-\delta H)$. This replacement results in an equivalent
Green's function to the order in $v$ to which we work. We implement it
because it ensures that the quarkonium spectra of the nrqcd scheme and
this second scheme, which we call the ``hybrid scheme'' are identical.
In both the nrqcd and hybrid schemes, we take $G_r(\bm{x},t)=0$ for $t <
0$. In each scheme, we define a corresponding ``advanced'' Green's
function $G_a$, which satisfies the same evolution equations as the
hermitian conjugate of $G_r$, but with the boundary condition that it
vanishes for $t > 0$. 

We note that the nrqcd and hybrid Green's functions correspond to the
same heavy-quark and antiquark propagation, except in the initial and
final time slices. The hybrid scheme applies $\delta H$ to every time
slice, including the initial and final time slices. In contrast, the
nrqcd scheme does not apply $\delta H$ to the initial and final time
slices, but applies $\delta H$ to all of the other time slices. In the
operator matrix elements that we measure, the four-fermion operator is
at the sink of the heavy-quark and antiquark propagators. Hence, in the
nrqcd scheme, the interactions in $\delta H$, which include those that
change the spin, are turned off for one time step on either side of the
four-fermion operator.

In order to see the effect of the nrqcd scheme (and related schemes
to be described later), let us initially ignore {\it all} of the
interactions of the heavy-quark and antiquark with the gauge field $A$,
including those in $H_0$ and $U_4$, in the time step on either side of
the four-fermion operator. We call this variant the nrqcdx scheme.
In the nrqcdx scheme, the vertices at which $A$ interacts with the
heavy-quark vanish on the time slice that contains the four-fermion
vertex. Since this condition is a restriction on the temporal distance
between the four-fermion vertex and the nearest heavy-quark-gluon
vertex, its effect can be incorporated into the heavy-quark propagator
that connects these vertices.\footnote{Note that additional vertices
along the heavy-quark Green's function cannot enter the time slice of
the four-fermion interaction because the heavy-quark propagators have a
definite time ordering.}  One simply sets this propagator to zero at
zero temporal separation. Now let us examine how this affects the
momentum-space Feynman rule for the propagator. The momentum-space
free-field heavy-quark propagator is given by the Fourier transform of
the temporal propagator $\exp[-{\bf k}^2t/(2m)]$, which has support only
for $t\geq 0$:
\begin{equation}                                                      
\tilde{G}_0(k_0,{\bf k}) = \sum_{t=0}^\infty e^{\mp i k_0 t}
                   \exp\left(-\frac{{\bf k}^2}{2m} t\right).
\end{equation}
Here $k$ is the heavy-quark or antiquark momentum, the upper (lower)
sign is for the quark (antiquark), and we have approximated the lattice
Hamiltonian by the continuum expression. In the nrqcdx scheme, the
vanishing of the propagator at zero temporal separation is implemented
by removing the $t=0$ contribution from the above sum. That is, one
replaces a free-field heavy-quark propagator adjacent to the
four-fermion vertex with
\begin{eqnarray}
\tilde{G}_1(k_0,{\bf k}) &=& \sum_{t=1}^\infty e^{\mp i k_0 t}
                \exp\left(-\frac{{\bf k}^2}{2m} t\right)   \nonumber \\
                &=& \exp\left(\mp i k_0 -\frac{{\bf k}^2}{2m}\right)
                            \,\tilde{G}_0(k_0,{\bf k}).
\label{eqn:cutoff}
\end{eqnarray}
On the right-hand side of Eq.~(\ref{eqn:cutoff}), the oscillatory factor
$\exp(\pm i k_0)$ provides additional damping of the $k_0$
integration, while the Gaussian factor $\exp[-{\bf k}^2/(2m)]$ cuts off
the integration over the spatial components of the momentum at
$|{\bf k}|\sim \sqrt{m}$. In the absence of these factors, the cutoff is
of the order of the maximum value of the components of the lattice
momentum, {\it i.e.}, $\pi$. In the nrqcdx scheme, the factors in
Eq.~(\ref{eqn:cutoff}) cut off diagrammatic loops that contain
interactions in $\delta H$, but only for those loops that involve the
four-fermion vertex. We note that these factors appear twice in each
loop---once for each of the two heavy-quark propagators in the loop that
attach to the four-fermion vertex.

As we have mentioned, the NRQCD matrix elements that we measure contain
contributions to that grow as powers of the lattice cutoff. These arise
in the lowest nontrivial order in perturbation theory from loops
involving the four-fermion vertex. Hence, the change in the
effective cutoff for such loops that is provided by the nrqcdx scheme
helps to control the numerical size of the power-behaved contributions
in our simulations. We emphasize that the change in the effective cutoff
affects only the interactions that renormalize  the four-fermion
operator. It has no effect, for example, on the interactions that
produce the quarkonium masses or wave functions.

It is useful to examine the effect of the nrqcdx scheme in coordinate
space. If the four-fermion operator is at $t=0$, then first interactions
after the four-fermion interaction do not occur until $t=1$, at which time
the heavy-quark Green's function is just the free-field Green's function
\begin{equation}
G_r(1,{\bf x}) = \left(\frac{m}{2\pi}\right)^{3/2} 
                       \exp\left(\frac{-m x^2}{2 } \right).
\label{smearing}
\end{equation}
Hence, in coordinate space, the effect of the nrqcdx scheme is to smear
out the four-fermion interaction spatially over a distance of order
$1/\sqrt{m}$. 

Now let us return to the effects of including, in the time slices that
are adjacent to the four-fermion interaction, the interactions with the
gauge field $A$ that are contained in $H_0$ and $U_4$. This discussion is
relevant to the nrqcd scheme and to the coulomb scheme, which we
introduce later. By ignoring the effects of the Coulomb gluon field
$A_0=\phi$, we have ignored the effects of the potential in the
Schr\"{o}dinger equation. This is a reasonable approximation until the
separation of the $Q\bar Q$ pair is of the order of the typical
$Q\bar{Q}$ separation in the bound state, at which point one can take
the spatial smearing to be given roughly by the size of the $Q\bar Q$
bound state. Since we work in the Coulomb gauge, the effects of the
heavy-quark interactions with the spatial gauge field ${\bf A}$ are
subleading in $v$ relative to the effects of the potential.

\section{Calculations and results}
\label{sec:calcs-results}
We are interested in matrix elements between quarkonium states of operators 
of the form
\begin{subequations}
\label{operator-defns1}
\begin{eqnarray}
{\cal O}_1\left({}^1S_{0,0}\right)&=&\psi^{\dag}\chi\chi^{\dag}\psi, \\
{\cal O}_1\left({}^3S_{1,\pm 1}\right)&=&\psi^{\dag}\sigma_\mp\chi 
\chi^{\dag}\sigma_\pm\psi,                       \\
{\cal O}_1\left({}^3S_{1,0}\right)&=&\psi^{\dag}\sigma_3\chi
\chi^{\dag}\sigma_3\psi,
\end{eqnarray}    
\end{subequations} 
and
\begin{subequations}
\label{operator-defns8}
\begin{eqnarray}
{\cal 
O}_8\left({}^1S_{0,0}\right)&=&\psi^{\dag}T^a\chi\chi^{\dag}T^a\psi, \\   
{\cal O}_8\left({}^3S_{1,\pm 1}\right)&=&\psi^{\dag}\sigma_\mp T^a\chi       
\chi^{\dag}\sigma_\pm T^a\psi,                       \\                       
{\cal O}_8\left({}^3S_{1,0}\right)&=&\psi^{\dag}\sigma_3 T^a\chi 
\chi^{\dag} \sigma_3 T^a\psi,
\end{eqnarray}
\end{subequations}
where the subscripts $1$ and $8$ indicate color-singlet and color-octet
operators, respectively, and $S$ denotes an $S$-wave operator. In
${}^{2s+1}S_{s,m}$, $s$ is the total spin quantum number, $m$ is the
quantum number of the component of the spin along the quantization axis,
$T^a$ is an ${\rm SU}(3)$ color matrix in the fundamental representation
satisfying ${\rm Tr}(T^aT^b)=(1/2)\delta^{ab}$, and the $\sigma$'s are Pauli
spin matrices. In Eqs.~(\ref{operator-defns8}), there is an implied sum
over the color index $a$. The matrix elements of the operators in
Eqs.~(\ref{operator-defns1}) and (\ref{operator-defns8}) in quarkonium
states are those that appear in the NRQCD factorization formula for
decays (\ref{decay-fact}). We note that, according to the $v$-scaling rules
of NRQCD, the matrix elements of the color-octet operators are
suppressed by at least $v^3$ relative to the leading color-singlet
matrix elements.

We are interested in the matrix elements of the operators in
Eqs.~(\ref{operator-defns1},\ref{operator-defns8}) between both the 
spin-singlet (pseudoscalar) and spin-triplet (vector) $S$-wave quarkonium 
states. What we measure is the ratio of octet to singlet matrix elements
\begin{equation}
R(s_i,m_i,s_f,m_f)=\frac{\langle {}^{2s_i+1}S_{s_i,m_i}|
{\cal O}_8\left({}^{2s_f+1}S_{s_f,m_f}\right)
|{}^{2s_i+1}S_{s_i,m_i}\rangle}
{\langle {}^{2s_i+1}S_{s_i,m_i}|{\cal O}_1\left({}^{2s_i+1}S_{s_i,m_i}\right)
|{}^{2s_i+1}S_{s_i,m_i}\rangle_{\rm VS}}.
\label{ratio}
\end{equation}
where the initial (final) spins and $z$-components of spin are $s_i$ ($s_f$)
and $m_i$ ($m_f$), respectively. The subscript VS indicates that we have
used the vacuum-saturation approximation \cite{Bodwin:1994jh} for the
denominator. That is, we replace the sum over intermediate states in
the operator with the vacuum state. We do, in fact, measure the
corresponding ratios for the case of color-singlet operators in the
numerator,  and we find that the spin-diagonal ratios are close to one,
indicating that our use of the vacuum-saturation approximation is valid and
that the off-diagonal matrix elements are all small, as is predicted by
$v$ scaling.

Now let us describe how we measure $R$ on the lattice. First, we
create a $Q\bar Q$ pair, using a source on a time slice $t$. We propagate
this pair to a second time slice $t' > t$, using the equations for the
retarded Green's functions given in Sec.~\ref{sec:NRQCD}. Then we
annihilate the $Q\bar Q$ pair at a point, in the spin and color state
that corresponds to the operator of interest. We then re-create the
$Q\bar Q$ pair at the same point and propagate this pair to time slice
$t'' > t'$, where it is annihilated at a sink. In practice we use sinks
on the time slice $t''$ as sources and propagate the $Q\bar Q$ pair back
to $t'$ by using the ``advanced'' Green's functions, conjugating these to
give us the Green's functions that we need. We call the quantity that we
have just described the ``lattice-matrix-element precursor.'' For
$T=t'-t$ and $T'=t''-t'$ sufficiently large, we annihilate the  $Q\bar
Q$ pair from an almost pure quarkonium ground-state wave function. The
quarkonium propagator falls as $\exp(-ET)$, where $E$ is the energy of
this ground state. By computing a ratio of color-octet to color-singlet
lattice-matrix-element precursors, we cancel this exponential falloff, as
well as the amplitude factors that are associated with the overlaps between
the sources and sinks and the quarkonium wave function. The result, in
the limit of large $T$ and $T'$, is precisely the lattice version of the
ratio $R$.

In our calculations, we use one of two stochastic sources on the initial
time slice to generate our retarded Green's functions. The first of these
sources consists of a complex random number that is uniformly
distributed in $U(1)$ at each site. The second of these sources consists
of a Gaussian smearing of this $U(1)$ random source. By choosing the $Q$
and $\bar Q$ Green's functions to have conjugate $U(1)$ sources, we
obtain a stochastic estimator of a point source for the $Q\bar{Q}$ pair
at each site on the initial time slice. By choosing the the $U(1)$ source
for the $Q$ and the conjugate of the Gaussian-smeared $U(1)$ source for
the $\bar Q$ (or {\it vice versa}), we obtain a stochastic estimator of a
Gaussian source for the $Q\bar{Q}$ pair at each site on the initial
time slice. The advanced Green's functions are treated similarly.

\begin{figure}[htb]
\vspace{-0.3in}
\epsfxsize=3.8in
\centerline{\epsffile{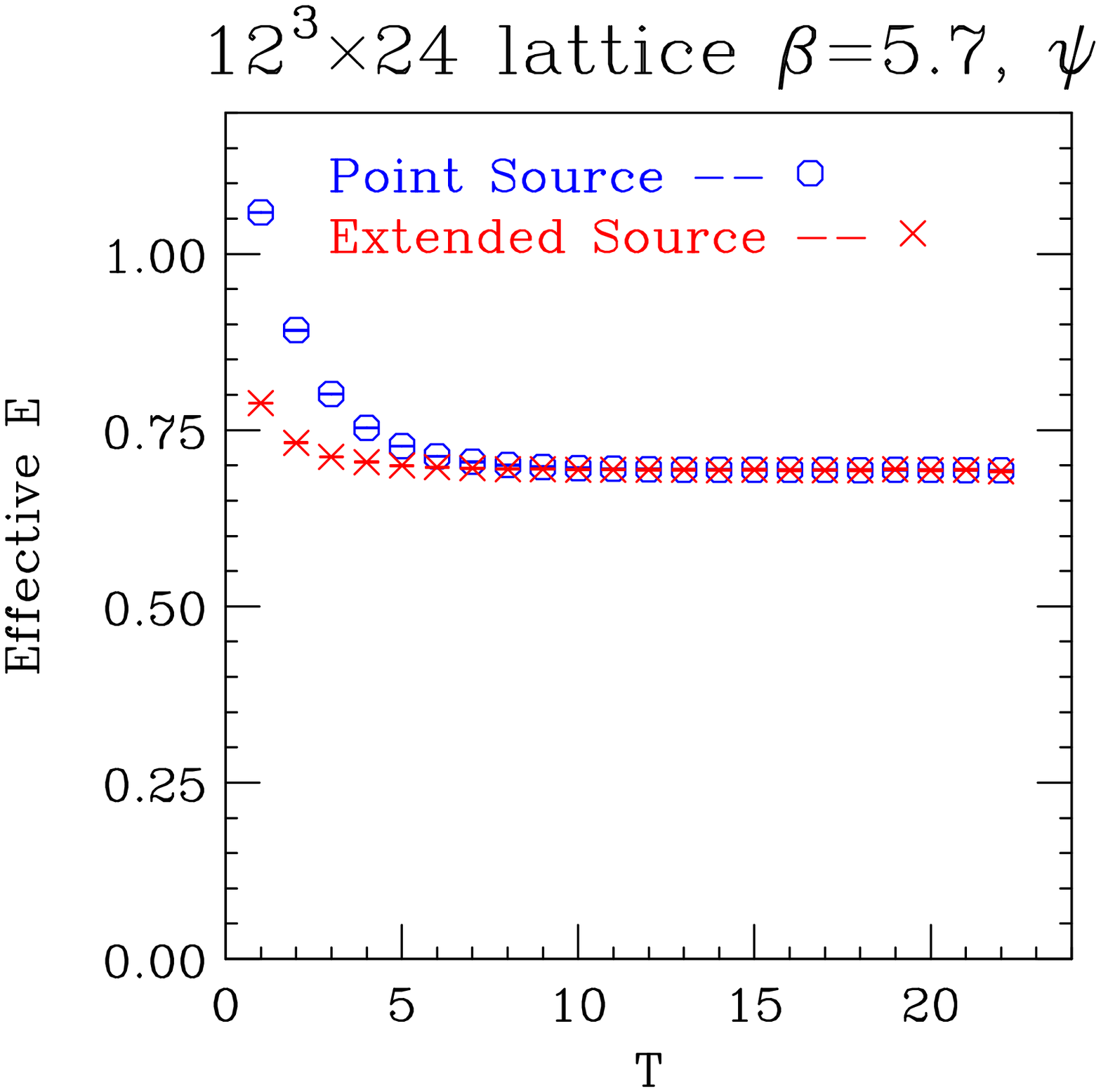}}
\vspace{0.2in}
\centerline{\epsffile{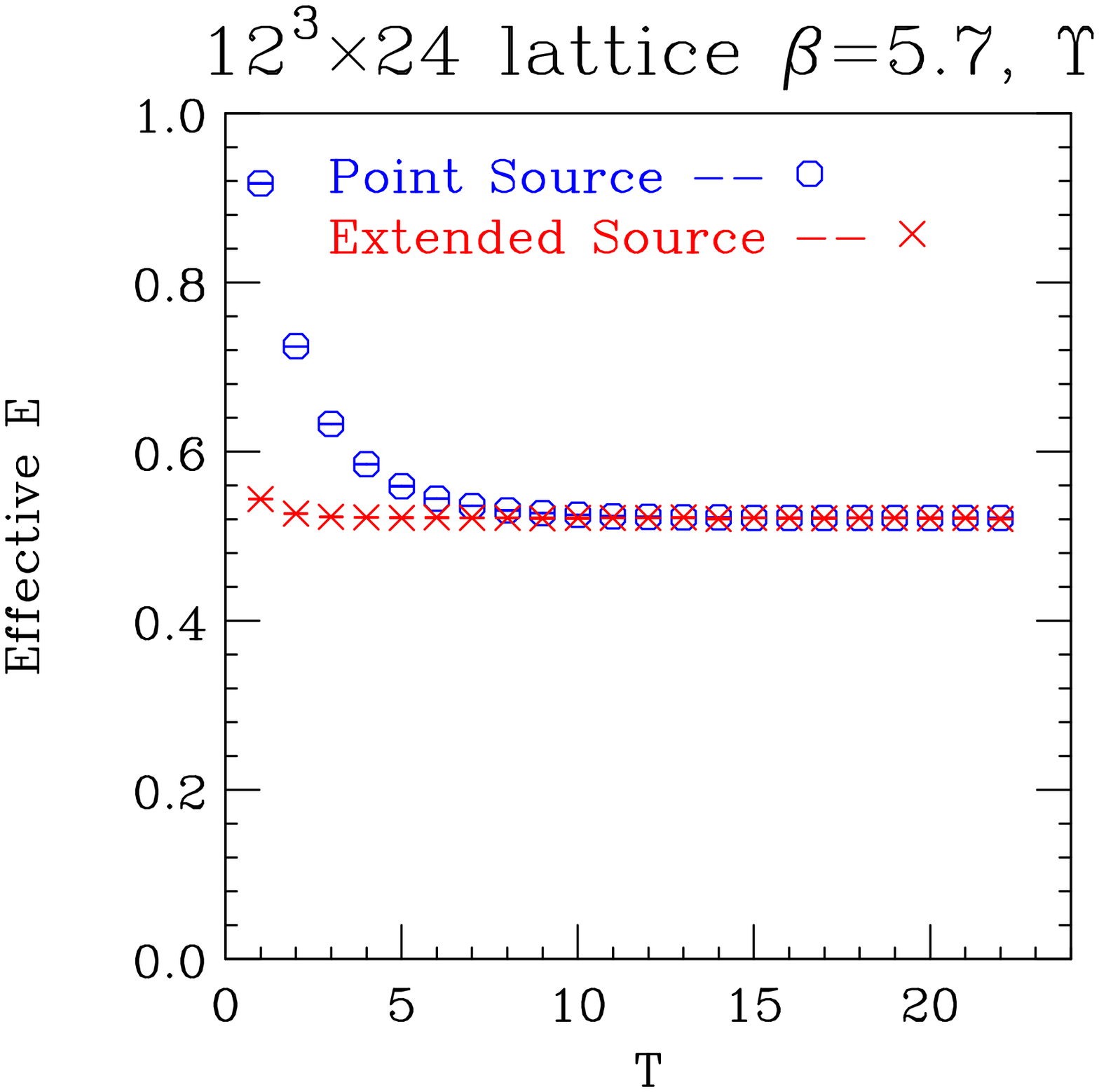}}
\caption{The effective quarkonium ``mass'' (energy) as measured with
point and Gaussian-smeared (extended) sources for the $J/\psi$ and the
$\Upsilon$. E is the effective energy, and T is the time separation
between the source and the sink.  The effective energies shown are
measured in the nrqcd updating scheme that is described in
Sec.~\ref{sec:NRQCD}. The plots in the hybrid scheme and for the
$\eta_c$ and the $\eta_b$ are similar.}\label{fig:mass}
\end{figure}

We calculate the quarkonium propagators and matrix elements on each of
400 equilibrated $12^3 \times 24$ quenched lattices at $\beta=5.7$. We use
the parameters that were determined by the NRQCD collaboration  
\cite{Davies:1994mp,Davies:1995db,Davies:1998im}. For $b$ quarks, we use
$m=3.15$, and for $c$ quarks we use $m=0.8$. We take
$u_0=0.860846184$, which we obtained from our own measurements of the 
average plaquette. We choose for the width of our Gaussian source 
$2.5$ lattice units, for both charmonium and bottomonium.
Fig.~\ref{fig:mass} shows effective quarkonium ``mass'' (energy) plots
for point and Gaussian-smeared sources as functions of the time
separation $T$ between the source and the sink. These plots indicate that
the Gaussian width that we take is reasonable, as it leads to an early
approach to the asymptotic value of the energy.
The fits to the effective energy in the nrqcd scheme with the extended
source/sink give an $\eta_c$ energy of $E_0=0.6165(6)$, with a
confidence level of 78\% for a fit over the $T(T')$-range 8--19, and a
$J/\psi$ energy of $E_1=0.6938(7)$, with a confidence level of 80\% for
a fit over the $T(T')$-range 10--22. The corresponding fits for bottomonium
give an $\eta_b$ energy of $E_0=0.5065(3)$, with a confidence level of
60\% over the $T(T')$-range 8--22, and an $\Upsilon$ energy of
$E_1=0.5215(3)$, with a confidence level of 59\% for a fit over the
$T(T')$-range 5--22. These results are to be compared with those of the 
NRQCD collaboration, which obtained $E_0=0.6182(7)$ for the $\eta_c$, 
$E_1=0.697(1)$ for the $J/\psi$, $E_0=0.5029(5)$ for the $\eta_b$, and
$E_1=0.5186(6)$ for the $\Upsilon$.

Now let us turn to the matrix elements. Here we recall that the decay
matrix elements that we measure are related to the production matrix
elements that are of primary interest through the 
crossing of the quarkonium from the intermediate state to the initial state.
In the production of the $J/\psi$ or the $\Upsilon$ by gluon
fragmentation---the dominant process at large $p_T$---the gluon fragments
into a $Q\bar Q$ pair in a triplet spin state with transverse
polarization. Hence, we are most interested in the decay matrix elements
in which the annihilated $Q\bar{Q}$ pair is transversely polarized.

We measure ratios $R$ of color-octet matrix elements to color-singlet
matrix elements defined in Eq.~(\ref{ratio}). For these measurements we
use both the hybrid and the nrqcd schemes for calculating the required
heavy-quark Green's functions.

Let us first examine the bottomonium matrix elements. Here the spin-singlet
quarkonium is the $\eta_b$ and the spin-triplet quarkonium is the $\Upsilon$.
We fit the measured ratios over a range of $T$ and $T'$ values with $T+T'<24$
in order to maximize the information that we extract from sometimes noisy 
``data.'' Our
measurements are shown in Table~\ref{tab:bottom}.
\begin{table}[htb]
\caption{Ratios $R$ of bottomonium color-octet matrix elements to
color-singlet matrix elements, as defined in Eq.~(\ref{ratio}). The
label ``singlet'' denotes a state ${}^1S_{0,0}$, ``triplet'' denotes a
state ${}^3S_{1,m}$, ``up'' denotes a state ${}^3S_{1,+1}$, ``down''
denotes a state ${}^3S_{1,-1}$, and ``longitudinal'' (long.) denotes a
state ${}^3S_{1,0}$. The spin transition $i\rightarrow f$ is from an
initial state $i$ to an annihilated state $f$ in the numerator of $R$.
In the case of triplet states, we average over $m_i$ in initial states
and sum over $m_f$ in annihilated states. The labels ``nrqcd'' and
``hybrid'' refer to the methods for calculating the heavy-quark lattice
NRQCD Green's functions. The column labeled ``$v$~scaling'' gives the
$v$-scaling factors of the various ratios $R$, along with their
numerical values. In evaluating the $v$-scaling factors, we take
$v^2=0.1$ for bottomonium.}
\begin{tabular}{c@{\hspace{0.2in}}r@{\hspace{0.2in}}r@{\hspace{0.2in}}r}
\hline
\multicolumn{1}{c}{spin transition} & \multicolumn{1}{c}{ hybrid}
           & \multicolumn{1}{c}{nrqcd} & \multicolumn{1}{c}{$v$~scaling}    \\
\hline
singlet $\rightarrow$ triplet & $7.169(6)\times 10^{-3}$
        & $2.72(4) \times 10^{-4}$ & $v^3/(2N_c) \approx 5.3 \times 10^{-3}$ \\
triplet $\rightarrow$ singlet & $2.414(3)\times 10^{-3}$
        & $9.0(1)  \times 10^{-5}$ & $v^3/(2N_c) \approx 5.3 \times 10^{-3}$ \\
singlet $\rightarrow$ singlet & $6.1(5) \times 10^{-5}$ 
        & $6.5(5) \times 10^{-5}$  & $v^4/(2N_c) \approx 1.7 \times 10^{-3}$ \\
triplet $\rightarrow$ triplet & $8.1(6) \times 10^{-5}$ 
        & $6.9(5) \times 10^{-5}$  & $v^4/(2N_c) \approx 1.7 \times 10^{-3}$ \\
up      $\rightarrow$ up      & $7.3(6) \times 10^{-5}$
        & $6.9(6) \times 10^{-5}$  & $v^4/(2N_c) \approx 1.7 \times 10^{-3}$ \\
long. $\rightarrow$ transverse& $2.7(3) \times 10^{-6}$
        & $1-2    \times 10^{-6}$  & $v^6/(2N_c) \approx 1.7 \times 10^{-4}$ \\
down    $\rightarrow$ up      & $5.53(2) \times 10^{-6}$
        & $< 5    \times 10^{-8}$  & $v^6/(2N_c) \approx 1.7 \times 10^{-4}$ \\
\hline
\end{tabular}
\label{tab:bottom}
\end{table}

The errors presented in Table~\ref{tab:bottom} are statistical only. For
the matrix elements that connect the singlet and triplet spin states
and, in the hybrid scheme,  for the up~$\rightarrow$~down matrix
element, the signal is excellent, and the systematic errors associated
with our choices of fits are probably less than the statistical errors.
Such estimates of fitting errors are obtained by examining the fits over
a number of choices of fitting ranges in $T$ and $T'$ (132 choices for
each matrix element) and taking into account the range of fitted
values and statistical errors for the matrix element of interest. Where
possible, we consider only those fits for which $T$ and $T'$ are large
enough that the value of the matrix element appears to have reached a
stable plateau. In the remainder of this paper, when we refer to
systematic errors, we mean systematic errors associated with our choices
of fits, unless we explicitly state otherwise.
 
For the three diagonal matrix elements
in both schemes, in which the signal is noisy but still substantial, the
systematic errors could be as large as 20\%. The signal for the
longitudinal~$\rightarrow$~transverse matrix element in the hybrid
scheme is sufficiently weak that the systematic errors could be 
as much as 50\%. Finally, the signals for the two triplet spin-flip matrix
elements in the nrqcd scheme are so noisy that we are only willing to
place bounds on their values.

Now let us compare our results with expectations from the $v$-scaling
rules of NRQCD. The $v$-scaling factors of the ratios $R$ are given in
the last column of Table~\ref{tab:bottom}. For purposes of using the
$v$-scaling factors to estimate the sizes of matrix elements, we include
the color factor $1/(2N_c)$ that arises in the free $Q\bar{Q}$ matrix
elements, as suggested in Ref.~\cite{Petrelli:1997ge}. We evaluate the
$v$-scaling factors by taking $v^2=0.1$ for bottomonium. Such
$v$-scaling estimates are based on the assumption that the coefficients
in the expansions of the matrix elements in powers of $v$ are of order
unity. It is that assumption that we wish to test by our explicit
calculation. Of course, if it turns out that the coefficients are
significantly greater than order unity, then the $v$ expansion will be
of little use.

As we have mentioned, the NRQCD matrix elements that we measure contain
contributions that diverge as powers of $\Lambda$ in the limit $\Lambda
\rightarrow \infty$. These power-divergent contributions potentially
violate the $v$-scaling rules, which were derived under the assumption
that $\Lambda$ is of order $mv$. For the particular matrix elements that
we measure, a perturbative analysis shows that the leading contributions
to the singlet$\rightarrow$~triplet matrix elements diverge as
$[\alpha_s(\Lambda)/\pi](\Lambda/m_b)^2$, the diagonal contributions
diverge as
$[\alpha_s(\Lambda)/\pi]^2(\Lambda/m_b)^2\log^2(\Lambda/m_b)$, and the
triplet spin-flip contributions diverge as
$[\alpha_s(\Lambda)/\pi]^2(\Lambda/m_b)^4$. Unless $\Lambda$ is not much
larger than $mv$, these power-behaved contributions may lead to
significant numerical violations of the $v$-scaling rules. As we
mentioned earlier, we wish to test not only whether the $v$-scaling rules
hold for cutoffs $\Lambda \sim m_bv$, but also whether they continue
to hold for cutoffs $\Lambda \sim m_b$. Therefore, in testing the
$v$-scaling rules in the bottomonium system, we choose $\Lambda/m_b 
\sim 1$. As we stated earlier, at $\beta=5.7$, the lattice
momentum cutoff itself is close to the input $b$-quark mass. Hence, in
the case of bottomonium, no special choice of heavy-quark Green's
functions is required in order to control power-behaved contributions,
and we focus on the hybrid scheme.

In the hybrid scheme, we note that the singlet~$\rightarrow$~triplet
matrix elements are the largest color-octet matrix elements, as is
predicted by the $v$-scaling estimates. The $v$-scaling estimates also
predict their magnitudes correctly. The heavy-quark spin symmetry of
NRQCD predicts that the singlet~$\rightarrow$~triplet matrix element
should be a factor $3$ larger than the triplet~$\rightarrow$~singlet
matrix element, with corrections to that relation of order $v^2$. This
prediction is borne out by our measurements. The diagonal matrix elements
are suppressed relative to the singlet~$\rightarrow$~triplet and
triplet~$\rightarrow$~singlet matrix elements, as expected, but by
considerably more than would be expected from the $v$-scaling factors
alone, suggesting that their coefficients in the $v$ expansion are
small. The triplet spin-flip matrix elements are smaller still, as is
expected from the $v$-scaling estimates, but, again, they are much
smaller than would be expected from the $v$-scaling factors alone. The
suppression of the triplet spin-flip matrix elements relative to the
diagonal matrix elements is approximately what would be expected from the
$v$-scaling factors.

The effects of a change in the effective cutoff $\Lambda$ can be seen
when we compare the hybrid-scheme and nrqcd-scheme values for $R$ in
Table~\ref{tab:bottom}. The nrqcd scheme results in a smaller effective
cutoff for the interactions in $\delta H$, which include all of the
spin-dependent interactions. We see that the nrqcd-scheme
singlet~$\rightarrow$~triplet and triplet~$\rightarrow$~singlet matrix
elements are suppressed by a factor of about $26$ relative to the
corresponding hybrid-scheme matrix elements. The triplet spin-flip
matrix elements also appear to be suppressed. The diagonal matrix
elements are virtually unchanged. That is expected because the leading
contribution to the diagonal matrix elements comes from $H_0$, whose
effective cutoff is not changed in going from the hybrid scheme to the
nrqcd scheme.

Now let us turn to the charmonium case. For charmonium, the
heavy-quark mass in lattice units is only 0.8, while, as usual, the components
of the lattice momentum can range up to $\pi$ in magnitude. Hence, we expect
that the lowering of the effective cutoff $\Lambda$ that is provided by
using the nrqcd scheme will be helpful in reducing the effects of the
power-behaved contributions. In the case of charmonium, we will also
consider an additional scheme for computing the heavy-quark Green's
functions in which the links $U_i(x)$ are set to unity on the time slice
that is associated with the four-fermion operator (the time slice in
which the $Q\bar Q$ pair is annihilated and re-created). In other
respects, this scheme, which we call the ``coulomb scheme'' is identical
to the nrqcd scheme. In the coulomb scheme, we are neglecting the
interactions of the heavy quark with the fields $A_i$ on the $Q\bar
Q$-annihilation time slice. Since we work in the Coulomb gauge, these
interactions are subleading in $v^2$. The effect of the coulomb scheme,
relative to the nrqcd scheme, is to lower the effective cutoff on $H_0$
so that it is the same as the effective cutoff on $\delta H$. Hence, we
expect the diagonal matrix elements to show the effects of a reduced
cutoff.

\begin{figure}[htb]
\epsfxsize=4in               
\centerline{\epsffile{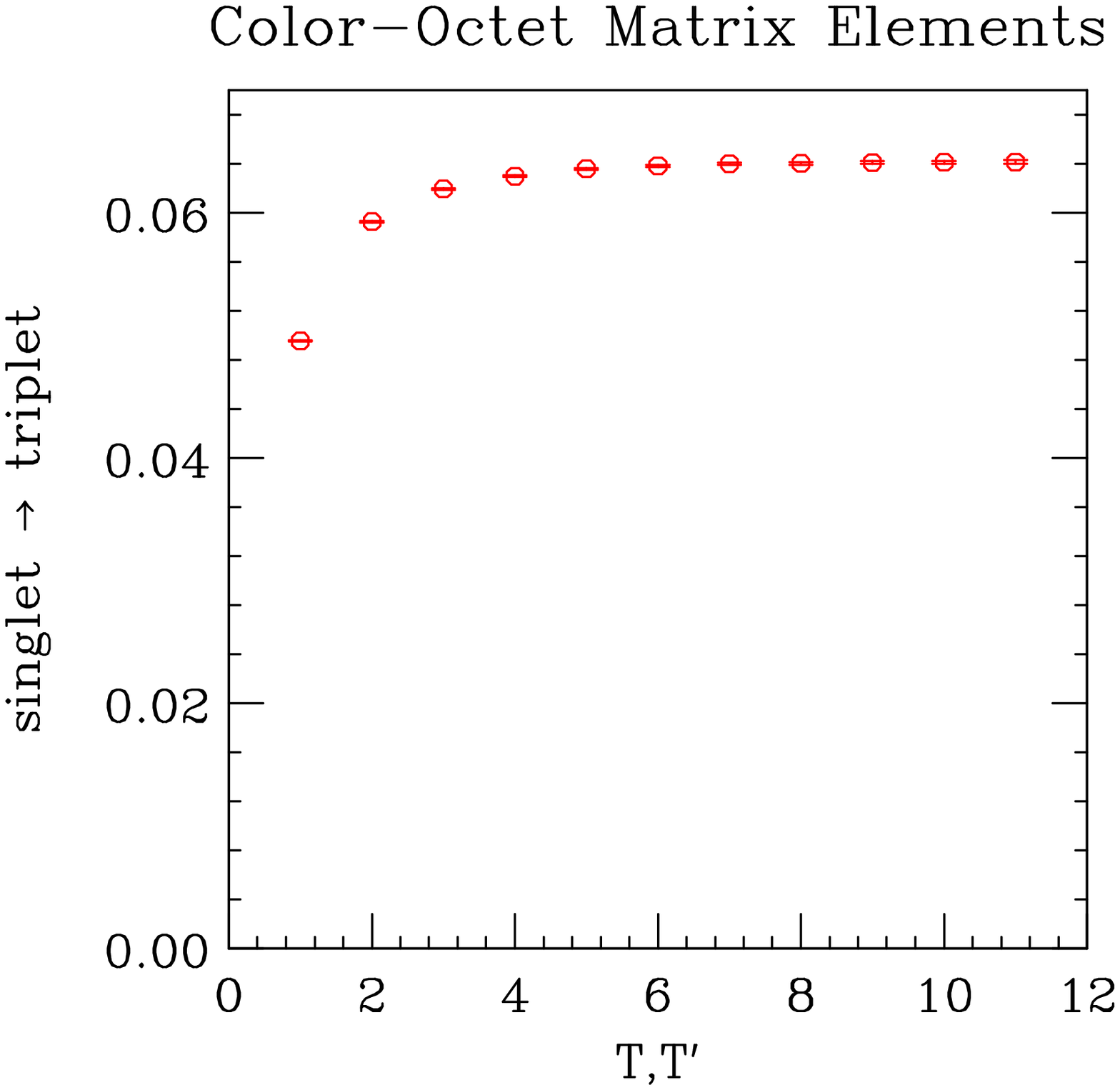}}
\vspace{0.2in}                                  
\centerline{\epsffile{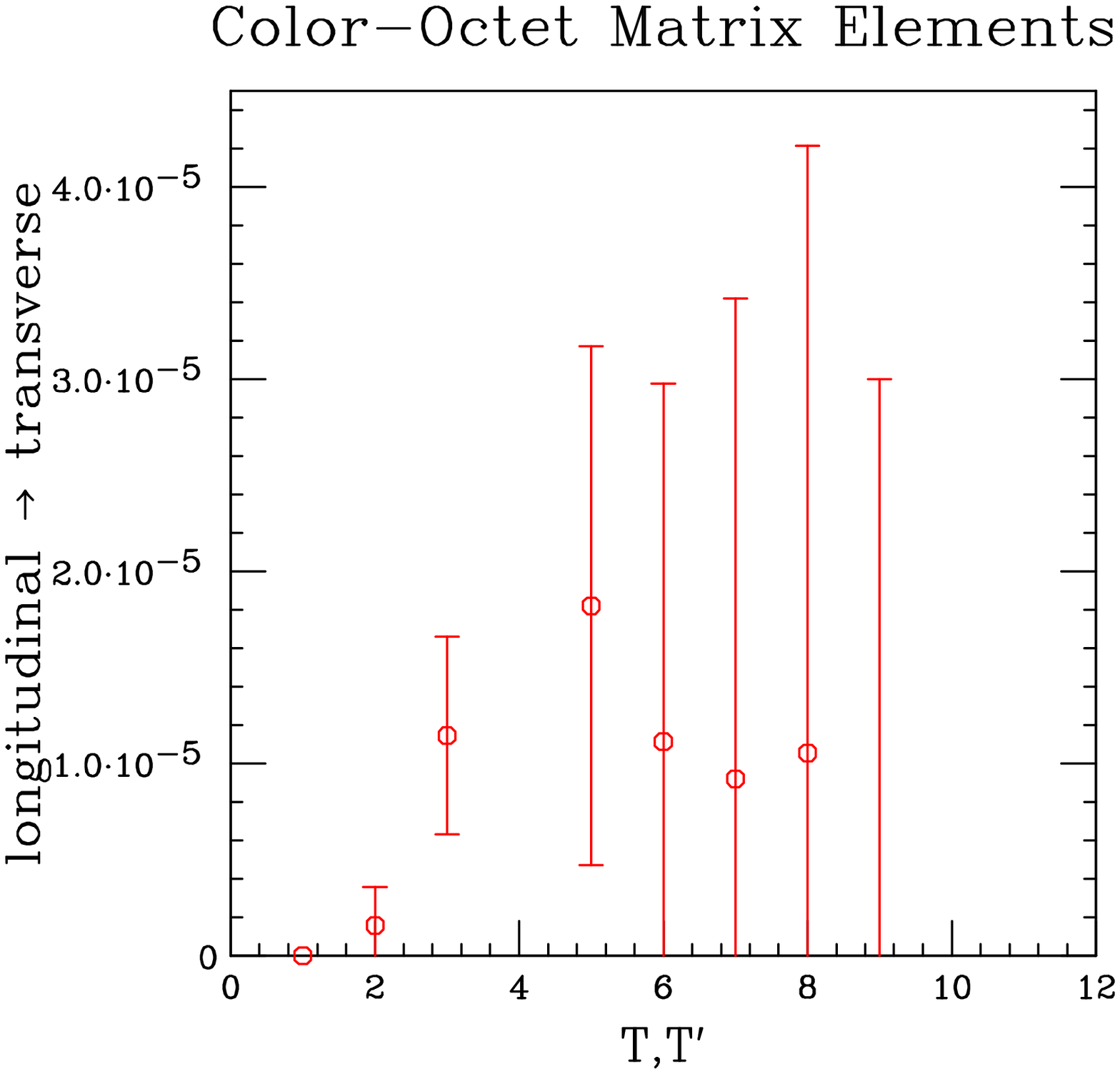}}
\caption{Values of ratios $R$ for charmonium as a function of $T=T'$.
The upper plot shows $R$ for the singlet~$\rightarrow$~triplet spin
transition in the hybrid scheme. The lower plot shows $R$ for the
longitudinal~$\rightarrow$~transverse spin transition in the coulomb
scheme.}\label{fig:emes}
\end{figure}

The values of $R$ in the hybrid, nrqcd, and coulomb schemes, along with 
the $v$-scaling factors, are given in Table~\ref{tab:charm}.
\begin{table}[htb]
\caption{Ratios $R$ of charmonium color-octet matrix elements to
color-singlet matrix elements, as defined in Eq.~(\ref{ratio}). The 
labels are as in Table~\ref{tab:bottom}, except that ``coulomb'' 
refers to an additional method for calculating the heavy-quark lattice NRQCD 
Green's functions, as described in the text. For purposes of numerical 
estimates of the $v$-scaling factors, we take $v^2=0.3$ for 
charmonium.}
\begin{tabular}{c@{\hspace{0.16in}}r@{\hspace{0.16in}}r@%
{\hspace{0.16in}}r@{\hspace{0.16in}}r}
\hline
\multicolumn{1}{c}{spin transition} & \multicolumn{1}{c}{ hybrid}
& \multicolumn{1}{c}{nrqcd} & \multicolumn{1}{c}{coulomb} &
\multicolumn{1}{c}{$v$~scaling}    \\
\hline
singlet $\rightarrow$ triplet &$6.397(8)\times 10^{-2}$&$2.90(3)\times 10^{-3}$
&$2.84(4)\times 10^{-3}$& $v^3/(2N_c) \approx 2.7\times 10^{-2}$ \\
triplet $\rightarrow$ singlet &$2.938(7)\times 10^{-2}$&$1.13(2)\times 10^{-3}$
&$1.06(1)\times 10^{-3}$& $v^3/(2N_c) \approx 2.7\times 10^{-2}$ \\
singlet $\rightarrow$ singlet &$5.03(9) \times 10^{-4}$&$9.7(2) \times 10^{-4}$
&$3.6(3) \times 10^{-4}$& $v^4/(2N_c) \approx 1.5\times 10^{-2}$ \\
triplet $\rightarrow$ triplet &$1.57(2)\times 10^{-3}$&$1.016(8)\times 10^{-3}$
&$4.0(4) \times 10^{-4}$& $v^4/(2N_c) \approx 1.5\times 10^{-2}$ \\
     up $\rightarrow$ up      &$4.57(6)\times 10^{-4}$&$1.019(8)\times 10^{-3}$
&$3.9(4) \times 10^{-4}$& $v^4/(2N_c) \approx 1.5\times 10^{-2}$ \\
long. $\rightarrow$ transverse&$2.82(4) \times 10^{-4}$&$2.8(7) \times 10^{-6}$
&\multicolumn{1}{c}{---}& $v^6/(2N_c) \approx 4.5\times 10^{-3}$ \\
        down $\rightarrow$ up &$8.48(5) \times 10^{-4}$&$1.4(2) \times 10^{-6}$
&$1.4(3) \times 10^{-6}$& $v^6/(2N_c) \approx 4.5\times 10^{-3}$ \\
\hline
\end{tabular}
\label{tab:charm}
\end{table}
Again, the quoted errors are statistical only. In the hybrid scheme, the
signal for the triplet~$\rightarrow$~singlet,
singlet~$\rightarrow$~triplet, and up~$\rightarrow$~down matrix elements
is robust, and we feel confident that the systematic errors are less than
the statistical errors. The signal for the singlet$\rightarrow$~singlet
matrix element is noisier, but still quite good, so that we feel that
the systematic error is probably no worse than the statistical 
error. The signals for the up~$\rightarrow$~up and
transverse~$\rightarrow$~longitudinal matrix elements are even noisier,
but still substantial. In these cases, the systematic errors 
might be as large as 10\%. In the nrqcd scheme, the signals for the
triplet~$\rightarrow$~singlet, singlet~$\rightarrow$~triplet, and
singlet~$\rightarrow$~singlet matrix elements are as clean as those in
the hybrid scheme. The signals for the two diagonal triplet matrix
elements show some rise as $T$ and $T'$ are increased, and the systematic
error could be as large as 10\%. The signals for the two triplet
spin-flip matrix elements are so noisy that the values given should be
considered only to be order-of-magnitude estimates. In the coulomb
scheme, the signals for all of the matrix elements are noisier than in
either of the other two schemes. The triplet~$\rightarrow$~singlet and
singlet~$\rightarrow$~triplet elements probably have systematic errors
that are comparable with their statistical errors. The diagonal matrix
elements could have systematic errors as large as 20\%, while the
signals for the triplet spin-flip matrix elements are so noisy
that we can only say that these matrix elements are small. In
Figure~\ref{fig:emes} we show examples of the signals for the charmonium
ratios $R$ as functions of $T=T'$.
The upper plot shows $R$ for the singlet~$\rightarrow$~triplet
transition in the hybrid scheme, which has one of the clearest signals
that we have seen. The lower plot shows $R$ for the
longitudinal~$\rightarrow$~transverse transition in the coulomb scheme,
which has one of the noisiest signals that we have seen.

Now let us compare these results with the predictions from the
$v$-scaling factors, which are shown in the last column of
Table~\ref{tab:charm}. As in the bottomonium case, we include the color
factor $1/(2N_c)$ that arises in the free $Q\bar Q$ matrix elements
(Ref.~\cite{Petrelli:1997ge}). The charmonium $v$-scaling factors are
the same as those for bottomonium, except that we evaluate them by setting
$v^2=0.3$. Again, as in the case of bottomonium, the
triplet~$\rightarrow$~singlet and singlet~$\rightarrow$~triplet matrix
elements in the hybrid scheme show good agreement with the $v$-scaling
predictions. The other matrix elements are somewhat smaller than would
be expected from the $v$-scaling predictions, perhaps indicating that
their coefficients in the $v$ expansion are small. However, here we
notice that the triplet spin-flip and non-spin-flip matrix elements are
of similar size. As we have mentioned, we know that the effective cutoff
in the charmonium case is larger than $m$ and that the spin-flip matrix
elements diverge as
$[\alpha_s(\Lambda)/\pi]^2(\Lambda/m_c)^4$ at large $\Lambda$, while the
non-spin-flip matrix elements diverge only as
$[\alpha_s(\Lambda)/\pi]^2(\Lambda/m_c)^2\log^2(\Lambda/m_c)$.
Therefore, it is not surprising that the triplet spin-flip matrix
elements appear to be anomalously large in comparison with the
non-spin-flip matrix elements.  In the nrqcd scheme, which effectively
lowers the cutoff on the interactions in $\delta H$, we see a
considerable reduction in the size of the triplet spin-flip matrix
elements, as well as in the triplet~$\rightarrow$~singlet and
singlet~$\rightarrow$~triplet matrix elements, which diverge as
$[\alpha_s(\Lambda)/\pi](\Lambda/m_c)^2$. However, the sizes of the
diagonal matrix elements are not reduced, but merely made closer in
value, as would be expected when all off-diagonal matrix elements are
small. This indicates that these diagonal matrix elements get sizable
contributions from the interactions in $H_0$, whose effective cutoff is
not reduced in the nrqcd scheme. In the coulomb scheme, in which the
effective cutoff of the interactions in $H_0$ is also reduced, we see
that the diagonal matrix elements are reduced in size relative to those
in the nrqcd scheme. The triplet~$\rightarrow$~singlet and
singlet~$\rightarrow$~triplet matrix elements are almost identical in
size to their nrqcd-scheme counterparts because the effective cutoff on
the interactions in $\delta H$ is the same (to leading order in $v$) in
both schemes. In addition, to the extent that we can infer anything from
the very noisy signals for the triplet spin-flip matrix elements in the
coulomb scheme, the values of these matrix elements are consistent with
their remaining unchanged in going from the nrqcd scheme to the coulomb
scheme. We can therefore conclude that, were we able to carry out the
lattice computation directly at $\Lambda \sim m_c$, the size ordering of
matrix elements would be as predicted by the $v$-scaling estimates.
However, the suppression of the matrix elements of higher order in $v$
would be even more than that which is predicted by those estimates,
indicating that the those matrix elements have relatively small
coefficients in the $v$~expansion. We conclude that the $v$~expansion is
a useful tool for establishing a hierarchy of charmonium decay matrix
elements.

\section{Summary and conclusions}
\label{sec:summary}

We have calculated, in lattice NRQCD, color-octet, spin-dependent matrix
elements that appear in the NRQCD factorization expression
(\ref{decay-fact}) for decays of bottomonium and charmonium. The lattice
action we used is correct through order $v^4$ and contains the
standard tadpole improvement, as well as order-$a^2$ improvements to the
terms of leading order in $v$. The decay matrix elements that we
calculated are related by crossing symmetry to the production matrix
elements that appear in the dominant contributions at large $p_T$ to the
production of the $\Upsilon$ and the $J/\psi$. In our calculations, we
made use of various forms of the lattice heavy-quark Green's functions,
which are equivalent to the order in $v$ to which we work, but which
allow us to vary the effective momentum cutoff for interactions of the 
heavy-quark fields with the gauge fields.
Our goals were to test the accuracy of estimates of the sizes of matrix 
elements that are based on the $v$-scaling rules of NRQCD and to test 
the convergence of the $v$~expansion. 

In the case of bottomonium, we found that estimates that are based on
the $v$-scaling factors are reasonable for the
singlet~$\rightarrow$~triplet and triplet~$\rightarrow$~singlet matrix
elements. The diagonal singlet~$\rightarrow$~singlet and
triplet~$\rightarrow$~triplet matrix elements are suppressed relative to
these leading-order matrix elements, but by even more than one would
expect from the $v$-scaling factors alone. The suppression of the
triplet spin-flip matrix elements relative to the triplet non-spin-flip
matrix elements is consistent with predictions that are based on the
$v$-scaling factors. In all cases, the $v$-scaling predictions correctly 
indicate the hierarchy of matrix elements.

In the case of charmonium, the situation is made less clear by the
fact that, in order to describe the quarkonium physics on the lattice,
we are forced to work with lattice spacings that correspond to UV
momentum cutoffs that are considerably larger than the heavy-quark mass.
Consequently, we expect the matrix-element calculations to contain large
contributions that grow as a power of the cutoff and that violate the
$v$-scaling rules of NRQCD. By using various forms of the heavy-quark
Green's functions, we are able to vary the effective UV cutoffs of the
matrix elements. The indications from our studies of the effects of
varying these cutoffs are that the $v$-scaling rules are a good guide as
to which matrix elements are important, but that the matrix elements are
smaller than one would expect from the $v$-scaling factors alone. Again,
the $v$-scaling predictions correctly indicate the hierarchy of matrix
elements. Of course, the cutoff dependences of the matrix elements are
canceled in physical quantities, such as the quarkonium production and
decay rates, by corresponding cutoff dependences in the accompanying
short-distance coefficients. A more complete understanding of the
effects of varying the UV cutoffs could be obtained by studying the
cutoff dependences of the matrix elements in perturbation theory. For
example, one could calculate the matching coefficients between lattice
matrix elements and continuum (dimensionally-regulated) matrix elements.
The matching coefficients are infrared safe and, hence, are amenable to a
perturbative treatment. Such a calculation involves both one- and two-loop
contributions in lattice and continuum perturbation theory, and is
beyond the scope of this paper.

Phenomenological color-octet production matrix elements for the
triplet~$\rightarrow$~triplet transition, normalized against the
color-singlet matrix elements, as in $R$ in Tables~\ref{tab:bottom} and
\ref{tab:charm}, lie in the ranges $5.1$--$16 \times 10^{-3}$ for the
$\Upsilon(1S)$ and $1.9$--$24 \times 10^{-3}$ for the $J/\psi$
(Ref.~\cite{Kramer:2001hh}). The lattice decay matrix elements for the
triplet~$\rightarrow$~triplet transition are somewhat smaller than the
phenomenological range of production matrix elements for the $J/\psi$
and considerably smaller than the phenomenological range of production
matrix elements for the $\Upsilon$. However, effects from multiple-gluon
radiation, which may decrease the phenomenological values of the
color-octet matrix elements, have not yet been included in the analyses
of the $\Upsilon$ matrix elements. Furthermore, decay matrix elements
need not be equal to the corresponding production matrix elements, even
though they have the same $v$-scaling behavior. We also note that
lattice matrix elements differ from continuum matrix elements. We do not
expect the differences to be large for tadpole-improved lattice
calculations, provided that the lattice cutoff is of order or smaller
than the heavy-quark mass. However, that situation does not hold in our
charmonium calculations.

The charmonium color-octet singlet~$\rightarrow$~triplet decay matrix
element is relatively large, which suggests that the color-octet
triplet~$\rightarrow$~singlet production matrix element may also be
large.  This would imply that, at large $p_T$ at the Tevatron, where the
color-octet, spin-triplet process dominates $S$-wave quarkonium
production, the $\eta_c$ production rate may be comparable to the
$J/\psi$ production rate. Once the size of the effective UV cutoff is
reduced, the longitudinal~$\rightarrow$~transverse decay matrix element
is small relative to the up~$\rightarrow$~up decay matrix element. This
suggests that a similar hierarchy may hold for the
transverse~$\rightarrow$~longitudinal and up~$\rightarrow$~up production
matrix elements, which would support the prediction of large
transverse polarization at large $p_T$ at the Tevatron.

\begin{acknowledgments}
We would like to thank Christine Davies, Peter Lepage, and
Junko Shigemitsu for helpful discussions on lattice NRQCD.
JL thanks the Argonne Theory Group for its hospitality. 
GTB and DKS are supported by the US Department of Energy under contract
W-31-109-ENG-38. JL is supported by a Korea Research Foundation
Grant (KRF-2004-015-C00092). 
\end{acknowledgments}


\begin{thebibliography}{999}

\bibitem{Braaten:1993rw}
  E.~Braaten and T.~C.~Yuan,
  Phys.\ Rev.\ Lett.\  {\bf 71}, 1673 (1993)
  [arXiv:hep-ph/9303205].

\bibitem{Braaten:1995cj}
  E.~Braaten and T.~C.~Yuan,
  Phys.\ Rev.\ D {\bf 52}, 6627 (1995)
  [arXiv:hep-ph/9507398].



\bibitem{Einhorn:1975ua}
M.~B.~Einhorn and S.~D.~Ellis,
Phys.\ Rev.\ D {\bf 12}, 2007 (1975).

\bibitem{Ellis:1976fj}
S.~D.~Ellis, M.~B.~Einhorn, and C.~Quigg,
Phys.\ Rev.\ Lett.\  {\bf 36}, 1263 (1976).

\bibitem{Carlson:1976cd}
C.~E.~Carlson and R.~Suaya,
Phys.\ Rev.\ D {\bf 14}, 3115 (1976).

\bibitem{Kuhn:1979kb}
J.~H.~K\"uhn,
Phys.\ Lett.\ B {\bf 89}, 385 (1980).

\bibitem{DeGrand:wf}
T.~A.~DeGrand and D.~Toussaint,
Phys.\ Lett.\ B {\bf 89}, 256 (1980).

\bibitem{Kuhn:1979zb}
J.~H.~K\"uhn, S.~Nussinov, and R.~R\"uckl,
Z.\ Phys.\ C {\bf 5}, 117 (1980) .

\bibitem{Wise:1979tp}
M.~B.~Wise,
Phys.\ Lett.\ B {\bf 89}, 229 (1980).

\bibitem{Chang:1979nn}
C.-H.~Chang,
Nucl.\ Phys.\ B {\bf 172}, 425 (1980).

\bibitem{Berger:1980ni}
E.~L.~Berger and D.~L.~Jones,
Phys.\ Rev.\ D {\bf 23}, 1521 (1981).

\bibitem{Baier:1981zz}
R.~Baier and R.~R\"uckl,
Nucl.\ Phys.\ B {\bf 201}, 1 (1982).

\bibitem{Baier:1981uk}
R.~Baier and R.~R\"uckl,
Phys.\ Lett.\ B {\bf 102} (1981) 364.

\bibitem{Keung:1981gs}
W.~Y.~Keung,
Print-81-0161 (BNL)
{\it Presented at Z0 Physics Workshop, Ithaca, N.Y., Feb 6-8, 1981}.

\bibitem{Baier:1983va}
R.~Baier and R.~R\"uckl,
Z.\ Phys.\ C {\bf 19} (1983) 251.


\bibitem{Brambilla:2004wf}
See 
N.~Brambilla {\it et al.},
arXiv:hep-ph/0412158.
and references therein.

\bibitem{Caswell:1985ui}
W.~E.~Caswell and G.~P.~Lepage,
Phys.\ Lett.\ B {\bf 167}, 437 (1986).
 
\bibitem{Thacker:1990bm}
B.~A.~Thacker and G.~P.~Lepage,
Phys.\ Rev.\ D {\bf 43}, 196 (1991).

\bibitem{Bodwin:1994jh}
G.~T.~Bodwin, E.~Braaten, and G.~P.~Lepage,
Phys.\ Rev.\ D {\bf 51}, 1125 (1995)
[Erratum-ibid.\ D {\bf 55}, 5853 (1997)]
[arXiv:hep-ph/9407339].

\bibitem{Lepage:1992tx}
G.~P.~Lepage, L.~Magnea, C.~Nakhleh, U.~Magnea, and K.~Hornbostel,
Phys.\ Rev.\ D {\bf 46}, 4052 (1992)
[arXiv:hep-lat/9205007].

\bibitem{Cho:1994ih}
P.~L.~Cho and M.~B.~Wise,
Phys.\ Lett.\ B {\bf 346}, 129 (1995)
[hep-ph/9411303].

\bibitem{Affolder:2000nn}
T.~Affolder {\it et al.}  [CDF Collaboration],
Phys.\ Rev.\ Lett.\  {\bf 85}, 2886 (2000)
[hep-ex/0004027].

\bibitem{Braaten:1999qk}
E.~Braaten, B.~A.~Kniehl, and J.~Lee,
Phys.\ Rev.\ D {\bf 62}, 094005 (2000)
[hep-ph/9911436].

\bibitem{Beneke:1997av}
M.~Beneke,
[hep-ph/9703429].
 
\bibitem{Brambilla:1999xf}
N.~Brambilla, A.~Pineda, J.~Soto, and A.~Vairo,
Nucl.\ Phys.\ B {\bf 566}, 275 (2000)
[hep-ph/9907240].
 
\bibitem{Fleming:2000ib}
S.~Fleming, I.~Z.~Rothstein, and A.~K.~Leibovich,
Phys.\ Rev.\ D {\bf 64}, 036002 (2001)
[hep-ph/0012062].

\bibitem{Sanchis-Lozano:2001rr}
M.~A.~Sanchis-Lozano,
Int.\ J.\ Mod.\ Phys.\ A {\bf 16}, 4189 (2001)
[hep-ph/0103140].
 
\bibitem{Brambilla:2002nu}
N.~Brambilla, D.~Eiras, A.~Pineda, J.~Soto, and A.~Vairo,
Phys.\ Rev.\ D {\bf 67}, 034018 (2003)
[hep-ph/0208019].

\bibitem{bks}
G.~T.~Bodwin, S.~Kim, and D.~K.~Sinclair,
Nucl.\ Phys.\ B (Proc.\ Suppl.)  {\bf 34}, 434 (1994);
{\bf 42}, 306 (1995)
[hep-lat/9412011];
G.~T.~Bodwin, D.~K.~Sinclair, and S.~Kim,
Phys.\ Rev.\ Lett.\  {\bf 77}, 2376 (1996)
[hep-lat/9605023];
Int.\ J.\ Mod.\ Phys.\ A {\bf 12}, 4019 (1997)
[hep-ph/9609371];
Phys.\ Rev.\ D {\bf 65}, 054504 (2002)
[hep-lat/0107011].

\bibitem{Davies:1994mp}
C.~T.~H.~Davies, K.~Hornbostel, A.~Langnau, G.~P.~Lepage, A.~Lidsey,
J.~Shigemitsu, and J.~H.~Sloan,
Phys.\ Rev.\ D {\bf 50}, 6963 (1994)
[arXiv:hep-lat/9406017].

\bibitem{Bodwin:2004up}
G.~T.~Bodwin, J.~Lee, and D.~K.~Sinclair,
in \textit{Quark Confinement and the Hadron Spectrum V},
AIP Conf. Proc. No. 756 (AIP, New York, 2005)
[arXiv:hep-lat/0412006].

\bibitem{Nayak:2005rw}
G.~C.~Nayak, J.~W.~Qiu, and G.~Sterman,
Phys.\ Lett.\ B {\bf 613}, 45 (2005)
[arXiv:hep-ph/0501235].

\bibitem{Lepage:1992xa}
G.~P.~Lepage and P.~B.~Mackenzie,
Phys.\ Rev.\ D {\bf 48}, 2250 (1993)
[arXiv:hep-lat/9209022].

\bibitem{Davies:1995db}
C.~T.~H.~Davies, K.~Hornbostel, G.~P.~Lepage, A.~J.~Lidsey, J.~Shigemitsu 
and, J.~H.~Sloan,
Phys.\ Rev.\ D {\bf 52}, 6519 (1995)
[arXiv:hep-lat/9506026].

\bibitem{Collins:2000ix}
S.~Collins, C.~T.~H.~Davies, J.~Hein, G.~P.~Lepage, C.~J.~Morningstar, 
J.~Shigemitsu, and J.~H.~Sloan,
Phys.\ Rev.\ D {\bf 63}, 034505 (2001)
[arXiv:hep-lat/0007016].

\bibitem{Davies:1998im}
C.~T.~H.~Davies, K.~Hornbostel, G.~P.~Lepage, A.~Lidsey, P.~McCallum, 
J.~Shigemitsu, and J.~H.~Sloan [UKQCD Collaboration],
Phys.\ Rev.\ D {\bf 58}, 054505 (1998)
[arXiv:hep-lat/9802024].

\bibitem{Petrelli:1997ge}
A.~Petrelli, M.~Cacciari, M.~Greco, F.~Maltoni, and M.~L.~Mangano,
Nucl.\ Phys.\ B {\bf 514}, 245 (1998)
[arXiv:hep-ph/9707223].

\bibitem{Kramer:2001hh}
M.~Kramer,
Prog.\ Part.\ Nucl.\ Phys.\  {\bf 47}, 141 (2001)
[arXiv:hep-ph/0106120].

\end{thebibliography}
\end{document}